\documentclass[aps,prl,twocolumn, superscriptaddress,amsmath]{revtex4} 
\usepackage[latin1]{inputenc}    
\usepackage{graphicx}   
\usepackage{xspace}

\begin{document}


\newcommand*{\siX}{\ensuremath{\sigma_\mathrm{x}}\xspace}
\newcommand*{\siZ}{\ensuremath{\sigma_\mathrm{z}}\xspace}

\newcommand*{\sz}{\ensuremath{|0\rangle}\xspace}
\newcommand*{\so}{\ensuremath{|1\rangle}\xspace}
\newcommand*{\PhiX}{\ensuremath{\Phi_\mathrm{X}}\xspace}
\newcommand*{\PhiZ}{\ensuremath{\Phi_\mathrm{Z}}\xspace}
\newcommand*{\mPhi}{\ensuremath{\mathrm{m\Phi_0}}\xspace}
\newcommand*{\fX}{\ensuremath{f_\mathrm{x}}\xspace}
\newcommand*{\fZ}{\ensuremath{f_\mathrm{z}}\xspace}
\newcommand*{\Ax}{\ensuremath{A_\mathrm{x}}\xspace}
\newcommand*{\Az}{\ensuremath{A_\mathrm{z}}\xspace}
\newcommand*{\czx}{\ensuremath{c_\mathrm{zx}}\xspace}

\newcommand*{\TF}{\ensuremath{T_{\varphi \mathrm{F}}}\xspace}
\newcommand*{\TE}{\ensuremath{T_{\varphi \mathrm{E}}}\xspace}
\newcommand*{\GF}{\ensuremath{\Gamma_{\varphi \mathrm{F}}}\xspace}
\newcommand*{\GE}{\ensuremath{\Gamma_{\varphi \mathrm{E}}}\xspace}

\newcommand*{\um}{\ensuremath{\,\mu\mathrm{m}}\xspace}
\newcommand*{\nm}{\ensuremath{\,\mathrm{nm}}\xspace}
\newcommand*{\mm}{\ensuremath{\,\mathrm{mm}}\xspace}
\newcommand*{\m}{\ensuremath{\,\mathrm{m}}\xspace}
\newcommand*{\sqm}{\ensuremath{\,\mathrm{m}^2}\xspace}
\newcommand*{\sqmm}{\ensuremath{\,\mathrm{mm}^2}\xspace}
\newcommand*{\squm}{\ensuremath{\,\mu\mathrm{m}^2}\xspace}
\newcommand*{\psqm}{\ensuremath{\,\mathrm{m}^{-2}}\xspace}
\newcommand*{\psqmV}{\ensuremath{\,\mathrm{m}^{-2}\mathrm{V}^{-1}}\xspace}
\newcommand*{\cm}{\ensuremath{\,\mathrm{cm}}\xspace}

\newcommand*{\nF}{\ensuremath{\,\mathrm{nF}}\xspace}
\newcommand*{\pF}{\ensuremath{\,\mathrm{pF}}\xspace}

\newcommand*{\emob}{\ensuremath{\,\mathrm{m}^2/\mathrm{V}\mathrm{s}}\xspace}
\newcommand*{\edos}{\ensuremath{\,\mu\mathrm{C}/\mathrm{cm}^2}\xspace}
\newcommand*{\mbar}{\ensuremath{\,\mathrm{mbar}}\xspace}

\newcommand*{\A}{\ensuremath{\,\mathrm{A}}\xspace}
\newcommand*{\mA}{\ensuremath{\,\mathrm{mA}}\xspace}
\newcommand*{\nA}{\ensuremath{\,\mathrm{nA}}\xspace}
\newcommand*{\pA}{\ensuremath{\,\mathrm{pA}}\xspace}
\newcommand*{\fA}{\ensuremath{\,\mathrm{fA}}\xspace}
\newcommand*{\uA}{\ensuremath{\,\mu\mathrm{A}}\xspace}

\newcommand*{\Ohm}{\ensuremath{\,\Omega}\xspace}
\newcommand*{\kOhm}{\ensuremath{\,\mathrm{k}\Omega}\xspace}
\newcommand*{\MOhm}{\ensuremath{\,\mathrm{M}\Omega}\xspace}
\newcommand*{\GOhm}{\ensuremath{\,\mathrm{G}\Omega}\xspace}

\newcommand*{\Hz}{\ensuremath{\,\mathrm{Hz}}\xspace}
\newcommand*{\kHz}{\ensuremath{\,\mathrm{kHz}}\xspace}
\newcommand*{\MHz}{\ensuremath{\,\mathrm{MHz}}\xspace}
\newcommand*{\GHz}{\ensuremath{\,\mathrm{GHz}}\xspace}
\newcommand*{\THz}{\ensuremath{\,\mathrm{THz}}\xspace}

\newcommand*{\K}{\ensuremath{\,\mathrm{K}}\xspace}
\newcommand*{\mK}{\ensuremath{\,\mathrm{mK}}\xspace}

\newcommand*{\kV}{\ensuremath{\,\mathrm{kV}}\xspace}
\newcommand*{\V}{\ensuremath{\,\mathrm{V}}\xspace}
\newcommand*{\mV}{\ensuremath{\,\mathrm{mV}}\xspace}
\newcommand*{\uV}{\ensuremath{\,\mu\mathrm{V}}\xspace}
\newcommand*{\nV}{\ensuremath{\,\mathrm{nV}}\xspace}

\newcommand*{\eV}{\ensuremath{\,\mathrm{eV}}\xspace}
\newcommand*{\meV}{\ensuremath{\,\mathrm{meV}}\xspace}
\newcommand*{\ueV}{\ensuremath{\,\mu\mathrm{eV}}\xspace}

\newcommand*{\T}{\ensuremath{\,\mathrm{T}}\xspace}
\newcommand*{\mT}{\ensuremath{\,\mathrm{mT}}\xspace}
\newcommand*{\uT}{\ensuremath{\,\mu\mathrm{T}}\xspace}

\newcommand*{\ms}{\ensuremath{\,\mathrm{ms}}\xspace}
\newcommand*{\s}{\ensuremath{\,\mathrm{s}}\xspace}
\newcommand*{\us}{\ensuremath{\,\mathrm{\mu s}}\xspace}
\newcommand*{\ns}{\ensuremath{\,\mathrm{ns}}\xspace}
\newcommand*{\rpm}{\ensuremath{\,\mathrm{rpm}}\xspace}
\newcommand*{\minute}{\ensuremath{\,\mathrm{min}}\xspace}
\newcommand*{\degree}{\ensuremath{\,^\circ\mathrm{C}}\xspace}

\newcommand*{\EqRef}[1]{Eq.~(\ref{#1})}
\newcommand*{\FigRef}[1]{Fig.~\ref{#1}}
\newcommand*{\dd}[2]{\mathrm{\partial}#1/\mathrm{\partial}#2}
\newcommand*{\ddf}[2]{\frac{\mathrm{\partial}#1}{\mathrm{\partial}#2}}

 \title{Time-Reversal Symmetry and Universal Conductance Fluctuations\\ in a Driven Two-Level System}
 \author{Simon Gustavsson}
 \affiliation {Research Laboratory of Electronics, Massachusetts Institute of Technology, Cambridge, MA 02139, USA}
 \author{Jonas Bylander}
 \affiliation {Research Laboratory of Electronics, Massachusetts Institute of Technology, Cambridge, MA 02139, USA}
 \author{William~D. Oliver}
 \affiliation {Research Laboratory of Electronics, Massachusetts Institute of Technology, Cambridge, MA 02139, USA}
 \affiliation {MIT Lincoln Laboratory, 244 Wood Street, Lexington, MA 02420, USA}

\begin{abstract}
In the presence of time-reversal symmetry, quantum interference gives strong corrections to the electric conductivity of disordered systems. The self-interference of an electron wavefunction traveling time-reversed paths leads to effects such as weak localization and universal conductance fluctuations. Here, we investigate the effects of broken time-reversal symmetry in a driven artificial two-level system. Using a superconducting flux qubit, we implement scattering events as multiple Landau-Zener transitions by driving the qubit periodically back and forth through an avoided crossing. Interference between different qubit trajectories give rise to a speckle pattern in the qubit transition rate, similar to the interference patterns created when coherent light is scattered off a disordered potential. Since the scattering events are imposed by the driving protocol, we can control the time-reversal symmetry of the system by making the drive waveform symmetric or asymmetric in time. We find that the fluctuations of the transition rate exhibit a sharp peak when the drive is time-symmetric, similar to universal conductance fluctuations in electronic transport through mesoscopic systems.
\end{abstract}

\maketitle



At low temperatures, the conductivity of disordered systems is strongly influenced by quantum interference effects such as weak localization (WL) and universal conductance fluctuations (UCF) \cite{Abrahams:1979,LeeRMP:1985,Altshuler:1988}.
Weak localization is due to constructive self-interference of an electron wavefunction traversing time-reversed paths, leading to an enhanced probability of back-scattering and therefore a reduction of the conductance.
UCF describe the strong fluctuations in conductance that occur as a function of any parameter that changes the scattering configuration \cite{Webb:1985,LeePRL:1985,Altshuler:1985}.
The interference effects are highly sensitive to anything that breaks time-reversal symmetry, such as a magnetic field applied perpendicularly to the motion of the charge carriers.
Studies of weak localization and UCF thus provide a method for investigating effects related to phase coherence, which has been used in a wide variety of systems ranging from metals \cite{Dolan:1979} and semiconductors \cite{Bishop:1980} to quantum dots \cite{Marcus:1992,Chan:1995,Folk:1996} and graphene \cite{Morozov:2006}, and even for the scattering of light off disordered media \cite{VanAlbadaMP:1985,Wolf:1985,Scheffold:1998}.

The presence of weak localization and UCF in such widely varying systems shows the universality of the effect, occurring independently of the sample size, dimensionality and the degree of disorder.
In mesoscopic systems, there are typically a large number of scatterers, giving millions of interfering paths that contribute to the electron transport.
In this work, we investigate the effect of time-reversal symmetry on a driven artificial two-level system in the few-scatterer limit.
The scattering events are implemented as Landau-Zener transitions by driving a qubit multiple times through an avoided crossing.
With both the number of scattering events and the time-reversal symmetry imposed by the driving protocol, we have control over the number of possible paths in the system.
In a configuration with only four scatterers, we measure a sharp increase in the fluctuations of the qubit transition rate when the drive waveform is made symmetric in time.

\begin{figure}[tb]
\centering
\includegraphics[width=\linewidth]{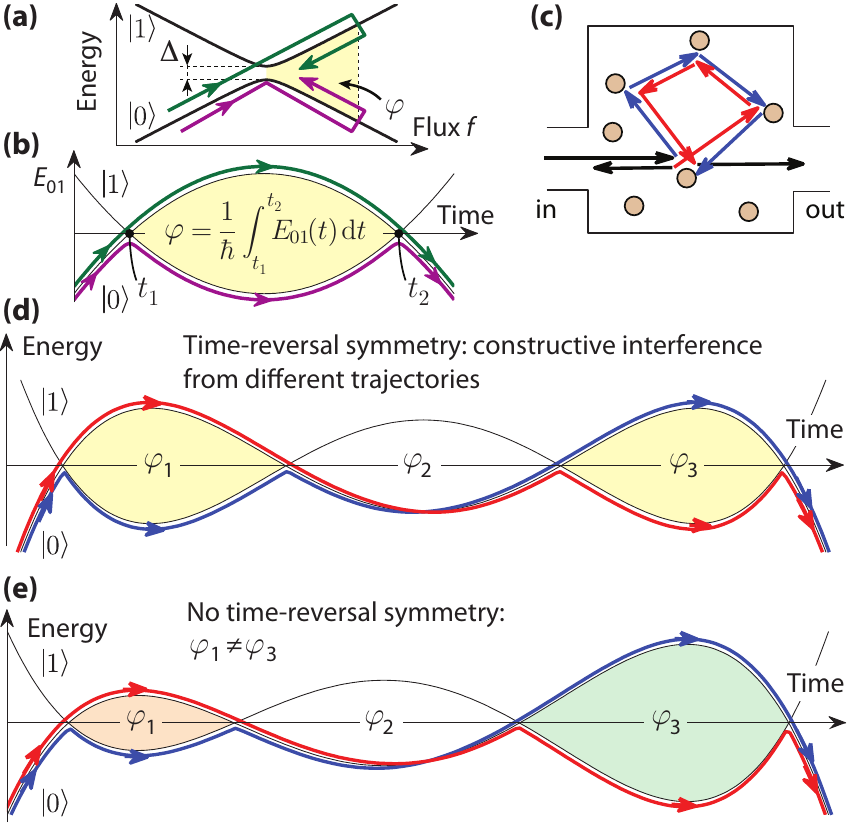}
\caption{Qubit trajectories during driven evolution. (a) Energy-level diagram of a flux qubit.  Driving the qubit through the avoided crossing induces Landau-Zener transitions between the two states. The states recombine when the qubit is brought back through the avoided crossing, with the final outcome depending on the phase accumulated during the flux excursion.
 (b) Qubit transitions visualized as two scattering events in a Mach-Zehnder interferometer setup.
 (c) Illustration of a mesoscopic system with two paths related by time-reversal symmetry.
 (d-e) Qubit transitions for a drive waveform that brings the qubit through the avoided crossing multiple times.  The blue and red traces mark two (out of eight) possible trajectories in the system. (d) When the waveform is symmetric in time, the trajectories acquire the same phases (since $\varphi_1=\varphi_3$), and will interfere constructively. (e) Without time-reversal symmetry, $\varphi_1\neq\varphi_3$.
} \label{fig:Setup}
\end{figure}

We use a superconducting flux qubit \cite{Mooij:1999,Orlando:1999}, which consists of a niobium loop interrupted by three Josephson junctions \cite{Oliver:2005}, with a magnetic flux $\Phi$ threading the loop.
%
The clockwise and counterclockwise persistent currents $\pm I_\mathrm{P}$, corresponding to the qubit's diabatic states, are tunnel coupled with strength $\Delta$.
The two-level Hamiltonian $H=(-1/2)(\Delta\siX + \varepsilon\siZ)$ describes the qubit dynamics, where
$\varepsilon = 2 I_\mathrm{P} f$, $f \equiv \Phi-\Phi_0/2$ is the flux detuning, $\Phi_0=h/2e$ the superconducting flux quantum, and $\siX$ and $\siZ$ are Pauli matrices.
The qubit energy separation $E_{01}= \sqrt{\varepsilon^2 + \Delta^2}$ is therefore controlled by the flux $\Phi$ in the loop [\FigRef{fig:Setup}(a)].
%

We first discuss the concept of scattering events in a driven two-level system.
Starting at negative flux detuning and with the qubit in its ground state, we apply a large-amplitude flux signal that drives the qubit through an avoided crossing and back again [\FigRef{fig:Setup}(a)].  At the first avoided crossing, the ground state $\sz$ undergoes a Landau-Zener transition and splits into a coherent superposition of $\sz$ and $\so$. The states evolve independently until the second time they reach the avoided crossing, where they interfere 
constructively or destructively depending on the relative phase $\varphi$ acquired between the two transitions.

The Landau-Zener transitions and the qubit evolution can be thought of as a phase-space analog of an optical Mach-Zehnder interferometer \cite{Oliver:2005}.  Figure \ref{fig:Setup}(b) shows the energy evolution of the qubit during the drive, where the interference phase $\varphi = (1/\hbar)\int_{t_1}^{t_2} {E_{01}}(t)\,\mathrm{d}t$ is given by the shaded area between the two scattering events.
The setup is similar to the mesoscopic system of \FigRef{fig:Setup}(c) in the limit of very few scatterers. The qubit phase space of \FigRef{fig:Setup}(b) contains only two scatterers, two possible trajectories and one interference phase.  The problem can be solved analytically, with the resulting qubit transition rate showing oscillations as a function of the interference phase $\varphi$ \cite{Berns:2006}.

We can increase the number of trajectories in our system by driving the qubit back and forth through the avoided crossing several times for each cycle of the driving waveform.
This allows us to increase the complexity of the system and move closer to the mesoscopic case while still having a controlled, non-chaotic phase space.
Figure \ref{fig:Setup}(d) shows the qubit being driven through the avoided crossing two additional times, giving a total of four scatters, three interference phases, and $2^3=8$ possible paths.
Note that the drive waveform is symmetric in time. Qubit trajectories that are related by the temporal symmetry will pick up the same phase during the driven evolution, and they will therefore interfere constructively.  Examples of time-reversed paths are plotted in red and blue in \FigRef{fig:Setup}(d).
We can easily break the time-reversal symmetry by making the drive waveform asymmetric in time, as shown in \FigRef{fig:Setup}(e):  the red and blue trajectories will acquire different phases since $\varphi_1$ and $\varphi_3$ are no longer equal.  
The qubit thus provides a well-controlled test system for investigating the effects of broken time-reversal symmetry in the few-scatterer limit.


\begin{figure}[tb]
\centering
\includegraphics[width=\linewidth]{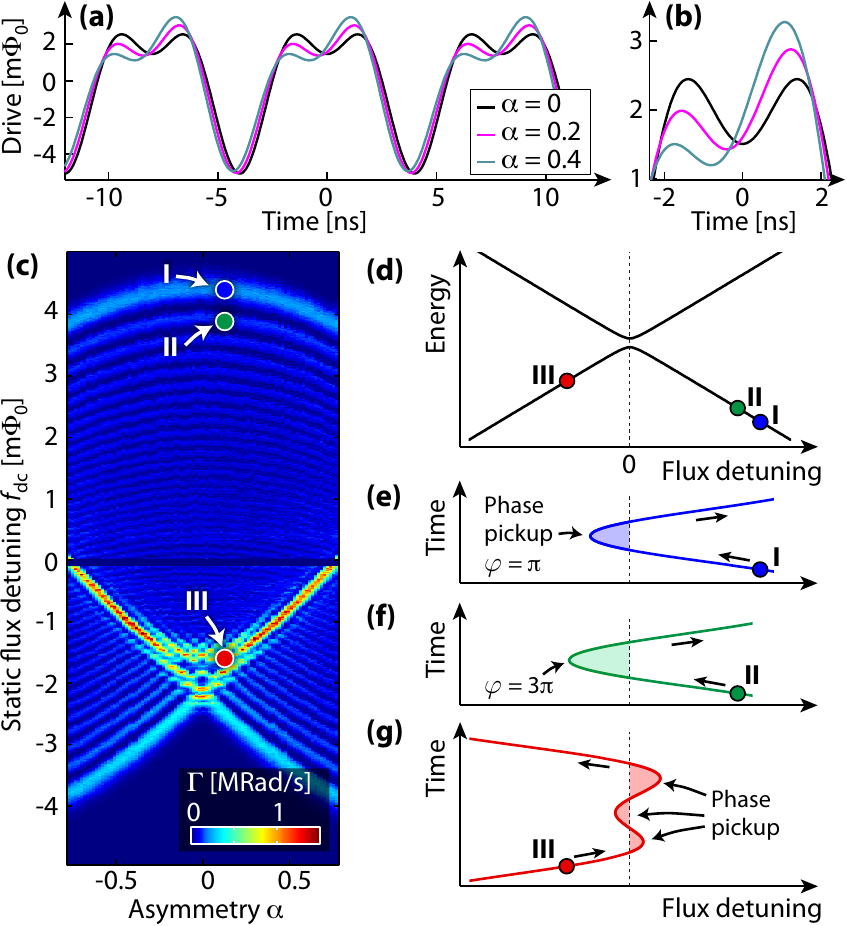}
\caption{Qubit transition rate as a function of the time-reversal symmetry of the drive waveform. (a) Drive waveform for different values of the drive asymmetry parameter $\alpha$.
 (b) Zoom-in around the center region of (a); the waveform becomes asymmetric in time when $\alpha \neq 0$.
 (c) Measured transition rate versus static flux detuning $f_\mathrm{dc}$ and $\alpha$.
 (d) Qubit energy diagram.
 (e-f) Case when $f_\mathrm{dc}>0$ (positions in (c) marked by I and II); the qubit is driven through the avoided crossing twice, resulting in Landau-Zener oscillations as a function of the accumulated phase $\varphi$.
 (g) Case when $f_\mathrm{dc}<0$ (position III); here, the qubit is driven through the avoided crossing four times, giving rise to an intricate interference pattern.
} \label{fig:Drive}
\end{figure}

We implement the drive protocol using a biharmonic signal \cite{Bylander:2009},
\begin{equation}
 f(t) = f_\mathrm{dc} + A_1 \cos(\omega t + \alpha) - A_2 \cos(2\omega t).
 \label{eq:Drive}
\end{equation}
We fix the frequency $\omega/(2\pi) = 125\MHz$, the amplitude $A_1 = 3\,\mPhi$ and the amplitude ratio $A_2/A_1=0.55$, while the parameter $\alpha$ controls the waveform's asymmetry. The waveform is plotted in \FigRef{fig:Drive}(a) for a few different values of $\alpha$, with \FigRef{fig:Drive}(b) showing a magnification around the time $t=0$.
Note that the function is symmetric in time for $\alpha=0$ and becomes increasingly asymmetric as $\alpha$ is increased.

The waveform's period $2\pi/\omega = 8\ns$ is comparable to the qubit's dephasing time $T_\phi \approx 10\ns$.  This puts the dynamics in the quasiclassical regime \cite{Berns:2006}, where coherence is preserved within one period of the drive, but where multi-photon processes due to coherence over many periods are not resolved \cite{Oliver:2005}. 
However, since the energy-relaxation time $T_1 \approx 20\us$ is much longer, consecutive periods of the drive signal will lead to a build-up of excited-state population.  The process of reaching the equilibrium population follows an exponential time dependence, which we characterize by a transition rate $\Gamma$.   In the relevant regime ($\Gamma \gg 1/T_1$), we have $\Gamma= 2W$, where $W$ is the probability (per unit time) of inducing a qubit transition from $|0\rangle$ to $|1\rangle$ after one period of the drive waveform \cite{Berns:2006}.

The measurement procedure consists of three steps:
first, the qubit is cooled to its ground state using a 3-$\mu$s cooling pulse \cite{Valenzuela:2006};
next, we apply the drive signal to induce qubit transitions;
finally, the qubit state is read out using a superconducting quantum interference device. 
%
By changing the length of the drive pulse and fitting the measured qubit population versus pulse duration to an exponential decay, we extract the transition rate $\Gamma$.
Figure \ref{fig:Drive}(c) shows a measurement of $\Gamma$ versus the static flux detuning $f_\mathrm{dc}$ and the asymmetry parameter $\alpha$.  The static component $f_\mathrm{dc}$ is used to shift the extrema of the waveform relative to the avoided crossing.  This is illustrated in Figs.~\ref{fig:Drive}(d-g), where we sketch the qubit energy bands and the flux excursion for three values of $f_\mathrm{dc}$.

For positive $f_\mathrm{dc}$, only the negative dip of the biharmonic signal reaches the avoided crossing.  This section of the wave will generate two Landau-Zener transitions just like in \FigRef{fig:Setup}(a), resulting in a transition rate $\Gamma$ that oscillates as a function of the interference phase $\varphi$ picked up between the scattering events [shaded area in Figs.~\ref{fig:Drive}(e)].  
At the maximum of $\Gamma$ marked by I in \FigRef{fig:Drive}(c), the drive just barely reaches the avoided crossing, giving a phase pickup of $\varphi=\pi$.
As $f_\mathrm{dc}$ is decreased, the flux excursion between scattering events gets longer and the phase pickup increases. The next maximum (point II) 
corresponds to $\varphi=3\pi$, and successive maxima occur whenever the interference phase equals odd integers of $\pi$. 
In this simplified picture, we do not consider the phase accumulation during the positive part of the flux sweep, which has a smaller influence on the resulting transition rate due to the qubit's comparatively short dephasing time ($T_\phi \approx 10\ns$). 

For negative $f_\mathrm{dc}$, the positive part of the biharmonic signal reaches the avoided crossing, as illustrated in \FigRef{fig:Drive}(g).
Depending on parameters, the waveform may drive the qubit through the avoided crossing up to four times per cycle, giving three interference phases. 
The phase accumulation and the interference conditions vary strongly with waveform shape, and this is the origin of the rich fluctuations in $\Gamma$ around point III in \FigRef{fig:Drive}(c).
Coming back to the trajectories discussed in Figs.~\ref{fig:Setup}(d-e), we can interpret the fluctuations in $\Gamma$ as interference from all possible paths generated by the scattering events.
Note that the drive waveform is periodic in time, so the number of possible scattering events is not restricted to the four events discussed in Figs.~\ref{fig:Setup}(d-e).  Rather, interference occurs between all trajectories that retain phase-coherence, and the maximal number of paths is ultimately set by the coherence time of the qubit.


%

\begin{figure}[tb]
\centering
\includegraphics[width=\linewidth]{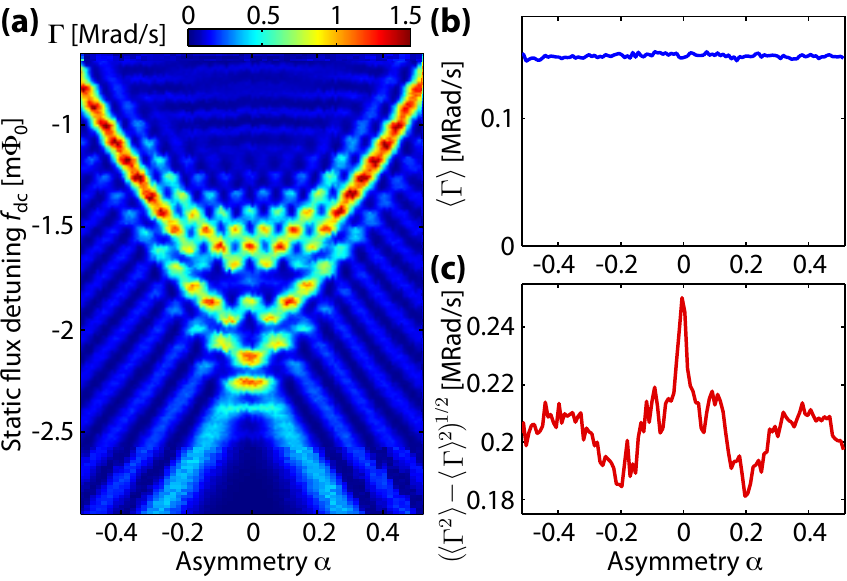}
\caption{Fluctuations in the transition rate of a driven qubit. (a) Measured transition rate vs flux detuning and drive asymmetry $\alpha$.
 (b) Transition rate averaged from $-4\,\mPhi$ to $0\,\mPhi$.  The data does not show any dependence on the drive asymmetry.
 (c) Standard deviation of the transition rate.  The fluctuations have a sharp peak at $\alpha = 0$.
} \label{fig:Rates}
\end{figure}

Figure \ref{fig:Rates}(a) shows a magnification of the region around point III in \FigRef{fig:Drive}(c). Despite involving only a few scattering events, the plot shows a rich interference pattern as a function of $f_\mathrm{dc}$ and $\alpha$.
To make the connection to charge transport in mesoscopic systems, we identify the qubit's transition rate $\Gamma$ with the electric conductance, whereas the time-symmetry breaking parameter $\alpha$ corresponds to the magnetic field. 
We calculate the mean and the standard deviation of $\Gamma$ by averaging over $f_\mathrm{dc}$ in the range $-4\,\mPhi < f_\mathrm{dc} < 0\,\mPhi$.
Since the parameter $f_\mathrm{dc}$ controls the timing of the scattering events, this averaging effectively corresponds to averaging over different scattering configurations.  In mesoscopic systems, the averaging over different scattering configurations is typically done by tuning a gate voltage or the in-plane magnetic field.


Figures \ref{fig:Rates}(b) and \ref{fig:Rates}(c) show the average transition rate $\langle \Gamma \rangle$ and the standard deviation of $\Gamma$, respectively, as a function of drive waveform asymmetry.
We note that $\langle \Gamma \rangle$ is independent of $\alpha$.  This is in contrast to mesoscopic systems, which typically show a dip in the conductance due to weak localization in the presence of time-reversal symmetry. 
Weak localization is thus not present in our experiments.  
On the other hand, the sharp peak in the standard deviation of $\Gamma$ at $\alpha = 0$ [\FigRef{fig:Rates}(c)] corresponds to the UCF-peak of phase-coherent electron transport.
The existence of UCF-like features without weak localization is related to our particular implementation of the driving protocol.  For a biharmonic drive signal, the average qubit transition rate is given by a sum of Bessel functions, which satisfy a sum rule \cite{Oliver:2005, Berns:2006} that holds regardless of waveform asymmetry.  
We calculate the $\alpha$-independent $\langle\Gamma\rangle = 0.15$\,Mrad/s
\footnote{To obtain the average transition rate, we integrate the perturbative expression for the transition rate $\Gamma\approx 2W$ in Ref.~\cite{Berns:2006} over flux detuning:
$\langle \Gamma\rangle =
(2\pi/2\Omega) \int_0^{f_\mathrm{dc}^\mathrm{max}} \mathrm{d}f_\mathrm{dc} \, \Gamma(f_\mathrm{dc}) =
(2\pi/2\Omega) \times \pi (\Delta/h)^2/2 = 0.15\,\mathrm{Mrad/s}$. The normalization $\Omega = 2m_0 |f_\mathrm{dc}^\mathrm{max}|$ includes the slope of the lowest energy level, $m_0 = \mathrm{d}E_0/\mathrm{d}f_\mathrm{dc} = 1.44\,\mathrm{GHz/m}\Phi_0$, and $f_\mathrm{dc}^\mathrm{max} = -4\,\mathrm{m}\Phi_0$. The UCF $\langle\Gamma^2\rangle$ does not have a simple analytic form.}, in agreement with the results of \FigRef{fig:Rates}(c).
We expect weak localization to be present when driving the qubit with a more arbitrary waveform, in which case the average transition rate will not be constrained by a sum rule.  
%

%

\begin{figure}[tb]
\centering
\includegraphics[width=\linewidth]{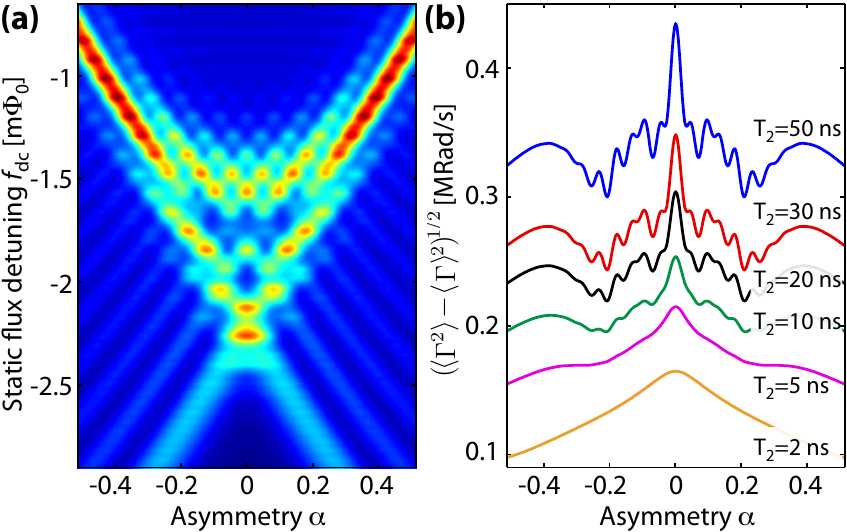}
\caption{Simulation results. (a) Simulated transition rate $\Gamma$ for qubit parameters $T_2 = 10\ns$ and $\Delta/h = 19\MHz$.
 (b) Standard deviation of the simulated transition rates for different values of $T_2$, extracted from simulation results similar to those shown in \FigRef{fig:Simulation}(a). Note that the curves are \emph{not} offset from each other.
} \label{fig:Simulation}
\end{figure}

Measurements of weak localization and universal conductance fluctuations are normally used to determine coherence lengths in two-dimensional systems.  For our driven qubit, where the interfering trajectories are in phase space rather than real space, we can employ the technique to investigate the coherence time $T_2 \,(\approx T_\phi)$ of the device.
To this end, we have simulated the qubit's transition rate numerically \cite{Berns:2006} for different values of $T_2$ under the drive defined by \EqRef{eq:Drive}.
Figure~\ref{fig:Simulation}(a) shows the result of such a simulation, with qubit parameters $\Delta/h = 19\MHz$ ($h$ is Planck's constant), $T_1 = 20\us$, and $T_2 = 10\ns$, in striking agreement with the measured data in \FigRef{fig:Rates}(a).

In \FigRef{fig:Simulation}(b), we plot the standard deviation of the transition rate, calculated by repeating the simulation for different values of $T_2$.  The peak around zero asymmetry is very broad for small values of $T_2$, but becomes sharper and stronger as $T_2$ is increased.  
Features due to weak localization and UCF in mesoscopic systems show the same behavior.
For long $T_2$, there are structures appearing in \FigRef{fig:Simulation}(b), away from the main peak at $\alpha = 0$, which are an effect of the small number of possible trajectories in our setup.  In mesoscopic systems, the large number of scattering possibilities average out all configuration-specific structures, and leaves only features related to the breaking of time-reversal symmetry.
We note that the simulations give good agreement with the data for $T_2 = 10-20\ns$, consistent with previous results \cite{Berns:2006}.

To conclude, we have investigated effects due to time-reversal symmetry in a driven two-level system where both the scattering events and the time-reversal symmetry are imposed by the driving protocol. This setup allows us to perform experiments in the regime of few scatterers, with a finite and controllable number of possible trajectories.  We find that effects similar to universal conductance fluctuations, normally associated with chaotic systems with a large number of scattering events, persist even in the few-scatterer limit.
In addition, our work shows an example of how a well-controlled, well-understood device like a qubit can be used to simulate more complex quantum systems.

\begin{acknowledgments}
We thank T. Orlando, M. Rudner and L. Levitov for helpful discussions.
\end{acknowledgments}

\bibliographystyle{apsrev}
\bibliography{DrivenUCF}

\end{document}